# The case against bimodal star formation in elliptical galaxies

B.K. Gibson[1,2,3]
[1] Mt. Stromlo & Siding Spring Observatories, Australian National University, Weston Creek, Australia ACT 2611
[2] Department of Astrophysics, University of Oxford, Keble Road, Oxford, UK OX1 3RH
[3] Department of Geophysics & Astronomy, University of British Columbia, Vancouver, British Columbia, Canada V6T 1Z4

**ABSTRACT**
We consider the present-day photometric and chemical properties of elliptical galaxies, adopting the bimodal star formation scenario of Elbaz, Arnaud & Vangioni-Flam (1995). These models utilise an initial mass function (IMF) biased heavily toward massive stars during the early phases of galactic evolution, leading to early Type II supernovae-driven galactic winds. A subsequent lengthy, milder, star formation phase with a normal IMF ensues, responsible supposedly for the stellar population observed today. Based upon chemical evolution arguments alone, this scenario has been invoked to explain the observed metal mass, and their abundance ratios, in the intracluster medium of galaxy clusters. Building upon the recent compilations of metallicity-dependent isochrones for simple stellar populations, we have constructed a coupled photometric and chemical evolution package for composite stellar populations in order to quantify the effects of such a model upon the photo-chemical properties of the resultant elliptical galaxies. We demonstrate that these predicted properties are incompatible with those observed at the current epoch.

## 1 INTRODUCTION

X-ray observations of the hot, optically thin, gas comprising the intergalactic medium of clusters of galaxies have established firmly the presence of a metal enriched component of non-primordial origin (e.g. Sarazin 1986 and Matteucci & Vettolani 1988, and references therein). Arnaud et al. (1992) have shown that a reasonably tight correlation exists between the total V-band luminosity of a cluster's elliptical-plus-lenticular population and the cluster's intracluster medium (ICM) iron mass – in particular, $M_{\rm Fe}^{\rm ICM} \approx 0.02 L_{\rm V}^{\rm E+S0}$. Determining $\alpha$-element to iron ratios for cluster ICMs is somewhat more problematic as the gas temperatures involved are usually too high to allow observation of emission lines besides iron. Recent ASCA measurements for the few clusters which have fairly secure oxygen and silicon detections do seem to indicate that an overabundance relative to iron exists, with [O/Fe] and [Si/Fe] ranging from $+0.20 \rightarrow +0.70$ and $+0.10 \rightarrow +0.60$, respectively (Mushotzky 1994).

Competing theories as to the origin of this metal enriched component of the ICM tend to fall into two camps – (i) ram pressure stripping of galactic interstellar media (ISMs) (Sarazin 1979), and (ii) supernovae (SNe)-driven winds subsequent to the initial phase of intense star formation in cluster ellipticals (Larson & Dinerstein 1975). Based upon cluster energetics and the lack of significant abundance gradients in the cores of the richest clusters, White (1991) has argued that the "stripping" scenario is not a viable one for most clusters. Additional evidence in support of the favoured "wind origin" for ICM metals can be seen in a number of recent papers, including Matteucci & Vettolani (1988), David, Forman & Jones (1991), Ciotti et al. (1991), Arnaud et al. (1992), Renzini et al. (1993), Mihara & Takahara (1994), Elbaz, Arnaud & Vangioni-Flam (1995), and Matteucci & Gibson (1995).

We note here, and simply refer to the aforementioned papers for further proof, that most of these wind models tend to underproduce iron relative to the observed relation of Arnaud et al. (1992) by factors of $\sim 3 \rightarrow 30$, *for models which honour <u>both</u> the Salpeter (1955) solar neighbourhood initial mass function (IMF) <u>and</u> the [$\alpha$/Fe]$\approx +0.4$ seen in the ICM.*

A typical "fix" for this situation has been to resort to an IMF for ellipticals which is biased toward massive stars (e.g. David, Forman & Jones 1991; Mihara & Takahara 1994; Matteucci & Gibson 1995), increasing the yield of Type II SNe-originating iron by suitably (and more im-portantly, freely) adjusting the IMF slope, star formation rate, and/or galactic wind epoch (i.e. the time at which the bulk of star formation is taken to cease) to match the ICM iron mass and $\alpha$-element ratios. Corroborating evidence for a flatter IMF in ellipticals usually references the population synthesis work of Arimoto & Yoshii (1987). As an aside, certain models either choose to relax, or were unaware of, the [O/Fe]>0 constraint (e.g. Matteucci & Vettolani 1988; Mihara & Takahara 1994) and will not be discussed further here.

An alternate fix to the iron underproduction "problem" retains the preferred Salpeter (1955) IMF for the bulk of the star formation phases, but postulates the existence of an earlier, more intense, stage of star formation (i.e. $10^5 \rightarrow 10^6$ M$_\odot$/yr with a Salpeter (1955) IMF truncated at a high lower mass limit of 3 M$_\odot$, reminiscent to that put forth for starburst galaxies (e.g. Charlot et al. (1993), and references therein). The initial star formation terminates after $\sim 20$ Myr due to the onset of a powerful Type II SNe-driven wind which expels, over the subsequent $\sim$40 Myr, $\sim$1/2 the initial galactic mass to the cluster ICM in the form of highly enriched gas. The second phase of star formation, with the "normal" IMF, then proceeds, the stars being formed out of the enriched material still present at the end of the wind phase. Again, by fine-tuning the star formation parameters and wind onset and duration, it is fairly straightforward to eject both the necessary amount of iron and in the proportions dictated by the ASCA observations. As pertains to elliptical galaxies and the ICM, this bimodal star formation model is introduced in the paper by Arnaud et al. (1992), and discussed in greater quantitative detail by Elbaz, Arnaud & Vangioni-Flam (1995). It is to the predictions of this latter work to which we will concentrate in the analysis which follows, although the conclusions are valid for the simple bimodal model as a whole.

The great danger in adopting either of the above "fixes" is that it treats cavalierly, or as is more usually the case, ignores entirely, the impact upon the resultant model ellipticals' photometric properties. It is of questionable use to construct synthetic elliptical models which explain the ICM abundances, but bear little resemblance to actual observed galaxies!

Unfortunately, until the very recent past it was extremely difficult to couple either of the above "chemical evolution motivated" models to photometric predictions due to the lack of non-solar metallicity stellar evolution tracks and photometric calibrations. It has only been with the release of Kurucz's (1993) model stellar atmospheres that a reasonably complete, and accurate, grid has become available for such an undertaking. The subsequent compilations of metallicity-dependent isochrones based upon the Kurucz grid, by Bertelli et al. (1994) and Worthey (1994), now make it possible to re-examine the photometric predictions for some of the galactic wind models adopted to explain the ICM abundances. To this end we have constructed a population synthesis package which builds upon the isochrones listed above and which is coupled to our chemical evolution code described elsewhere (Gibson 1995).

The bimodal star formation/IMF models mentioned earlier are successful at reproducing many of the *chemical* properties of elliptical galaxies. Because of their extreme iron production during the early phase of star formation (and its subsequent ejection to the ICM) in their models (e.g. ten to twenty times greater than the Case A models of Mihara & Takahara (1994)), and the above "favourable" chemical predictions, we felt it imperative to ascertain if the model was equally successful in replicating the present-day photo-chemical properties of the underlying ellipticals. *Self-consistent modeling of the photo-chemical properties of bimodal models has never been demonstrated before.* We stress that our goal is not to claim to have solved the so-called iron underproduction problem (although, see Gibson & Matteucci (1995) for the latest approach using the more conventional single IMF approach), but only to draw attention to shortcomings in their proposed bimodal star formation/galactic wind solution.

Section 2 introduces the photometric evolution technique adopted, and highlights some subtle differences between the Worthey (1994) and Bertelli et al. (1994) metal-rich isochrones. Missing low-mass extensions to the latter set are also described. The relevant equations in the Elbaz, Arnaud & Vangioni-Flam (1995) bimodal star formation scenario are then introduced, and the colours and metallicities of their favoured models presented in Section 3. Their predicted colour-luminosity-metallicity relation is then compared with observations, and the results summarised.

## 2  ANALYSIS

### 2.1  Isochrone synthesis

As the basis for our photometric properties analysis, we build upon the metallicity-sensitive isochrones of Worthey (1994) and Bertelli et al. (1994). General photometric evolution techniques and history are discussed at length in these two papers, and the numerous references therein. Both compilations span $\sim$2.5 dex in [Z] ([Z]$\equiv$log Z - log Z$_\odot$) and sample the primary stellar evolutionary stages from the zero age main sequence to well down the white dwarf cooling curve (or carbon ignition, depending upon the initial mass). Worthey (1994) does not include any post-asymptotic giant branch contribution in his tables, but comparisons done using Bertelli et al. (1994), with and without the planetary nebulae/white dwarf stages, showed this to be an unimportant omission for the analysis which follows.

As the Bertelli et al. (1994) isochrones have a lower limit of 0.6 M$_\odot$, it was necessary to extend these down to $\sim$0.1 M$_\odot$, in order to include the contribution from M-dwarfs, and fully sample the full range of masses in the IMF. To this end, we transformed the newly available low temperature stellar models of Allard & Hauschildt (1995), and in particular, their log $g \equiv$ +5.0 grid, from the theoretical plane in which they are provided, to the observer's plane, via the mass-luminosity-temperature relationships shown in Figure 1. These were derived from the older models of VandenBerg et al. (1983), so while not ideally matched to the Allard & Hauschildt (1995) models, they were adequate for our purposes. Interpolation and extrapolation for the Bertelli et al. (1994) isochrone metallicities was done where necessary, and continuity at $m \approx 0.6$ M$_\odot$ in the $M_{Bol}$-T$_{eff}$ and $M_V$-colour planes between the two sources was ensured. As in Bruzual & Charlot (1993), the colour properties of these low temperature extensions were assumed to be unevolving.

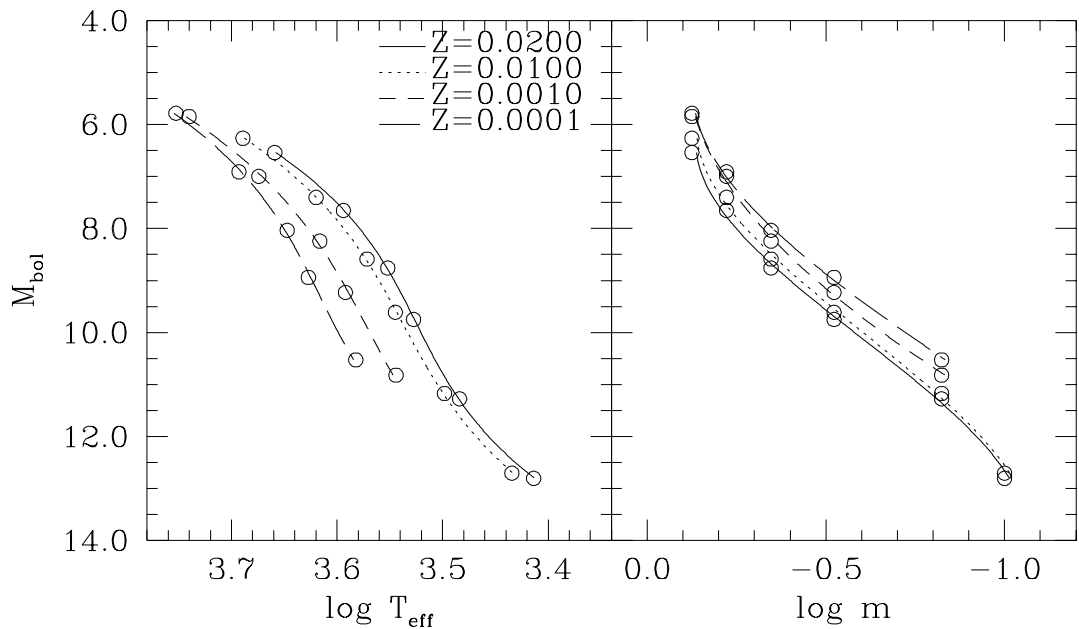

**Figure 1.** Effective temperature-mass-luminosity relations used to transform Allard & Hauschildt (1995) low temperature models to the observer's plane.

Worthey (1994), on the other hand, used the M-giant fluxes of Bessell et al. (1991), supplemented with optical spectrophotometry from the Gunn & Stryker (1983) stellar atlas, to extend his isochrones down to $m \approx 0.1$ $M_\odot$. Worthey's (1994) low temperature extension is virtually identical to that adopted above for the Bertelli et al. (1994) isochrones.

Table 1 lists the integrated V-K colours for simple stellar populations (SSPs) using the various isochrones of both Bertelli et al. (1994) and Worthey (1994). We have considered other colours in the analysis which follows, but we will stress V-K, as it provides a more valuable constraint than, say, B-V. Over the range of luminosities considered in our study, B-V does not vary by more than $\sim 0.2$ magnitudes, whereas the V-K versus $M_V$ relation is considerably flatter and spans $\gtrsim 1$ magnitude. As can be seen in Figure 7b of Arimoto & Yoshii (1987), the near-vertical distribution of ellipticals in the B-V versus $M_V$ plane makes this colour, by itself, a poorer constraint.

We do not wish to enter into a detailed analysis of the differences between the two sets of isochrones. Work of this nature can be found in Worthey's (1994) paper, and the important study by Charlot, Worthey & Bressan (1995), intercomparing the models of Worthey (1994), Bressan, Chiosi & Fagotto (1994), and Bruzual & Charlot (1993).

The one point which we do want to highlight in Table 1 is the behaviour of the super-solar colours. It is apparent that for metallicities $Z \lesssim Z_\odot$ the colours agree to within a few hundredths of a magnitude. This is clearly not the case for the $Z \approx 0.05$ isochrones, with Worthey's (1994) being $\sim 0.5$ magnitude redder in V-K. For metal-rich populations such as giant ellipticals, it is imperative to be aware of, and understand the origin of, this discrepancy.

**Table 1.** V-K colours for simple stellar populations of age $t_G \equiv 12$ Gyr and metallicity Z. BBCFN94=Bertelli et al. (1994)+low mass extensions; W94=Worthey (1994); *=Worthey (1995).

| Z | V-K BBCFN94 | W94 |
|---|---|---|
| 0.0950 | – | 4.62* |
| 0.0534 | – | 4.18 |
| 0.0500 | 3.68 | – |
| 0.0434 | – | 4.10 |
| 0.0361 | – | 4.02 |
| 0.0301 | – | 3.93 |
| 0.0200 | 3.39 | – |
| 0.0169 | – | 3.46 |
| 0.0095 | – | 3.08 |
| 0.0080 | 3.02 | – |
| 0.0053 | – | 2.78 |
| 0.0040 | 2.75 | – |
| 0.0017 | – | 2.42 |
| 0.0010 | 2.49 | – |
| 0.0005 | – | 2.23 |
| 0.0004 | 2.16 | – |
| 0.0002 | – | 2.18 |

Overlaying the Z=0.02 and Z=0.05 isochrones (only up to the red giant branch tip, for clarity), we see the difference quite graphically in Figure 2. The main sequences, sub-giant branches, and lower red giant branches agree for Z=0.05, but the upper giant branches, as seen in this $M_{bol}$–(V-K) plane, increasingly diverge. This divergence is not evident in the Z=0.02 isochrones, shown for comparison in the same plot.

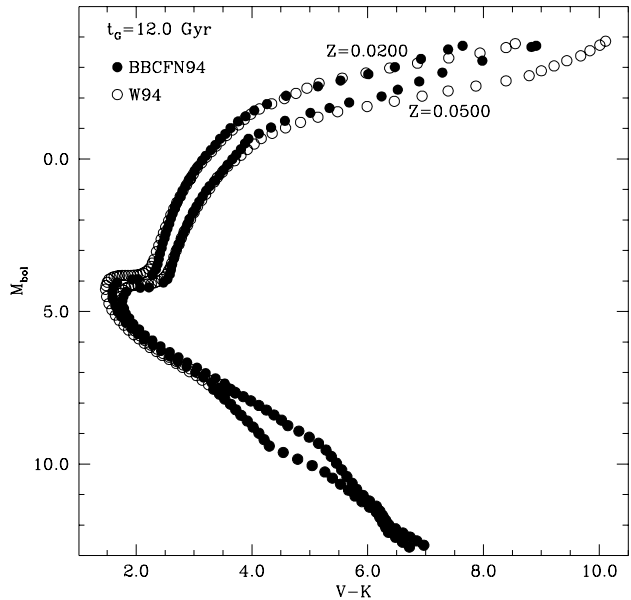

**Figure 2.** Solar and supersolar metallicity isochrones of age 12 Gyr. The post-red giant branch components have been eliminated for clarity. W94=Worthey 1994; BBCFN94=Bertelli et al. 1994.

The origin of this 0.5 magnitude discrepancy for Z≈ 0.05 seems to lie, at least in part, with the colour (V-K) – effective temperature relation used by the different groups. Whereas Worthey (1994) uses a metallicity-dependent scale for $Z>Z_\odot$ fluxes, Bertelli et al. (1994) do not (see their Sections 4(ii)-4(iv)). In Figure 3 we show the main sequence and red giant branch (V-K)–log $T_{eff}$ relations used for the solar (solid curves) and Z=0.05 (open and filled circles) isochrones from the two sources. Worthey's (1994) Z=0.05 giant branch can be seen to be $\sim 0.3 \rightarrow 0.6$ magnitudes redder in V-K, for a given temperature, when compared with his Z=0.02 giant branch. In contrast, the Bertelli et al. (1994) Z=0.05 giant branch relation overlays precisely their Z=0.02 relation, a partial contributor to their significantly bluer giant branch, when compared with Worthey (1994), for supersolar metallicities. We should stress that the gross offset in the RGB (V-K)–$T_{eff}$ plane for $Z\lesssim Z_\odot$, between Worthey (1994) and Bertelli et al. (1994), is masked/compensated by the two groups' different post-main sequence lifetimes and GB tips. This "conspiracy" means that the $Z\lesssim Z_\odot$ SSP integrated colours are always within $\sim 0.1$ mag (in V-K) of each other. Assuming this conspiracy holds for the Z=0.05 isochrones, which admittedly may not be entirely correct, implies that if Bertelli et al. (1994) scaled their GB redward, by what appears to be the more appropriate $\sim 0.3 \rightarrow 0.6$ mag (V-K), then the "usual" $\lesssim 0.1$ mag difference in the groups' colours would be recovered. It appears that Bressan, Chiosi & Fagotto (1994), who themselves use the Bertelli et al. (1994) isochrones, seem to have modified the latter's Z=0.05 (V-K)–log $T_{eff}$ relation, as their SSP integrated V-K is $\sim 4.3$, as opposed to the $\sim 3.7$ found in the digital version of the isochrones (at $t_G = 12$ Gyr). This is $\sim 0.15$ magnitudes redder than Worthey's (1994) equivalent isochrone (Table 1, and Worthey's (1994) Figure 35), but does appear to bring their late-time integrated Z=0.05 SSP colours more in line with those of Worthey (1994). This is *not* to imply that Worthey's (1994) approach is the "last word" on the subject; indeed, we are of the opinion that the stellar evolution models adopted by Bertelli et al. (1994) (and the subsequent Padua Group compilations) are superior to those used by Worthey (1994) (in particular, the reduced post-main sequence lifetimes in the Padua compilations seem preferable – Charlot, Worthey & Bressan 1995; Worthey 1995). Conversely, though, we believe that Worthey's (1994) spectral library (specifically, at low temperature, where his detailed colour-temperature analysis is most commendable), is preferred over that used in the Bertelli et al. (1994) isochrones.

The situation becomes further complicated when we examine the updated Bertelli et al. (1994) and Bressan, Chiosi & Fagotto (1994) isochrones (see Table 2 of Tantalo et al. 1995). Tantalo et al. (1995) have updated their (V-K)–log $T_{eff}$ relation, such that it now resembles very closely that used by Worthey (1994) (recall Figure 3). Intuitively, following the argument of the previous paragraph, this should lead to (V-K) colours which are $\sim 0.5$ redder than the Bertelli et al. (1994) isochrones, at late-time. (Aside: We speculate that this is indeed what was done by Bressan, Chiosi & Fagotto (1994), as it meshes with the colours quoted in their Table 3). Surprisingly, though, the (V-K) colours in Tantalo et al. (1995) are actually $\sim 0.1$ mag bluer than Bertelli et al. (1994) for Z=0.02, and roughly unchanged for Z=0.05. Again, we can only speculate, but we feel that the possible reason for this is that Tantalo et al. (1995) adopt an increased post-main sequence mass loss parametrisation ($\eta = 0.45$, as opposed to the conventional $\eta = 0.35$ used by Bressan, Chiosi & Fagotto 1994), which could contribute to their bluer SSPs through the consequent reduction in the RBG and AGB lifetimes.

Disentangling *all* the differences between the Bertelli et al. (1994), Bressan, Chiosi & Fagotto (1994), Tantalo et al. (1995), and Worthey (1994) isochrones is a difficult task. A very important first step in this process can be found in the excellent study by Charlot, Worthey & Bressan (1995) into some of the uncertainties in modeling stellar populations, to which the reader is directed for more details. We simply wish to illustrate what appears to be a key difference between the $Z>Z_\odot$ isochrones that was encountered, independently, during our analysis, in order to understand the differences which appear during the calculation of the present-day photo-chemical properties of our model ellipticals.

In what follows, we will provide parallel photometric evolution analyses based upon both the Bertelli et al. (1994) and Worthey (1994) isochrones. As to which set is "preferable" is left to the reader's discretion.

As a final note to implementing the Bertelli et al. (1994) isochrones, there is an error in the running tabulation of the infrared colours in the digital version of their tables, which will be corrected in future releases (Bertelli 1994).

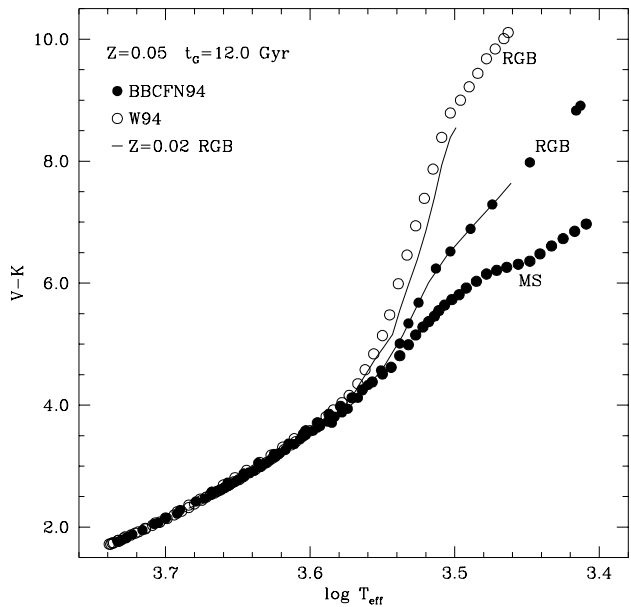

**Figure 3.** Colour-effective temperature relationship adopted for the solar and supersolar metallicity isochrones of Figure 2. Only the main sequence (MS) and red giant branch (RGB) are shown. Other symbols as in Figure 2.

### 2.2 Photometric evolution

Consider, first, the flux $F$ through the bandpass $\Delta\lambda$ for a composite stellar population (CSP) of metallicity $Z_V$ (given by equation 2), at time $t_G$, as the sum of a series of $i$ simple stellar populations (SSPs), each of age $t_G - t_i$ and metallicity $Z(t_i)$. For each SSP isochrone, we sum over the $n_{G-i}$ initial mass bins, each $j^{\rm th}$ bin having a luminosity $L_j$, resulting in an integrated flux

$$F^{\Delta\lambda}(t_G) = \sum_{i=0}^{t_{GW}/\Delta t} \sum_{j=1}^{n_{G-i}} \int_{m_{j,i,\ell}}^{m_{j,i,u}} \frac{\phi(m_{j,i})}{m_{j,i}} dm \, \psi(t_i) \, \Delta t \, L^{\Delta\lambda}_{j,G-i,Z(t_i)}. \quad (1)$$

The subscript "G-$i$" indicates the nearest isochrone of age $t_G - t_i$ to the metallicity $Z(t_i)$. $\phi(m) \propto m^{-x}$ is the initial mass function, by mass, where a slope $x = 1.35$ represents the canonical Salpeter (1955) value. $\psi(t)$ is the star formation rate.

Worthey (1994) ($Z \leq 0.05$) and Worthey (1995) ($Z > 0.05$) isochrones were kindly provided by Guy Worthey with a temporal resolution of 0.1 Gyr. $\sim$0.5 Gyr is the typical spacing for the Bertelli et al. (1994) tables. This was found to be more than adequate for our purposes. We are concerned predominantly with isochrones of age 10 $\rightarrow$ 12 Gyr for this work, as we will be comparing photo-chemical predictions against those seen in the present-day local elliptical population. The earliest phases of spectral evolution will not covered in this work.

Following Arimoto & Yoshii (1986), we assume that stars born in the discrete interval $\left[\sqrt{Z_{k-1}Z_k} \rightarrow \sqrt{Z_k Z_{k+1}}\right]$ are tied to the isochrone of metallicity $Z_k$. For the Worthey (1994) isochrones, the subscript $k$ runs from one to eleven, with $Z_1$=0.0002, $Z_2$=0.0005, $Z_3$=0.0017, $Z_4$=0.0053, $Z_5$=0.0095, $Z_6$=0.0169, $Z_7$=0.0301, $Z_8$=0.0361, $Z_9$=0.0434, $Z_{10}$=0.0534, and $Z_{11}$=0.0950. The first and last intervals are Z=0 $\rightarrow$ 0.00030 and Z=0.07127 $\rightarrow \infty$, respectively. For the Bertelli et al. (1994) isochrones, the subscript $i$ runs from one to six, with $Z_1$=0.0004, $Z_2$=0.0010, $Z_3$=0.0040, $Z_4$=0.0080, $Z_5$=0.0200, and $Z_6$=0.0500, with first and last intervals of Z=0 $\rightarrow$ 0.00063 and Z=0.03162 $\rightarrow \infty$, respectively.

We calculate a V-band luminosity-weighted metallicity which in spirit is similar to Arimoto & Yoshii (1987), but differs in that we calculate the mean metallicity <Z>, whereas in their original work they determined < [Z] > (i.e. they replaced Z with [Z] in equation 2). On the surface this would seem to be a trivial difference, but as it turns out, this has crucial implications for estimating the metallicity for the model galaxy. Using the strict Arimoto & Yoshii (1987) formalism results in any low metallicity component being weighted far too heavily. A simple "back-of-the-envelope" example which illustrates this point might be useful – let us take a simple scenario whereby we have ten stars of equal luminosity, one star of which has Z=0.02 (i.e. [Z]=+0.000) and the other nine have Z=0.00002 (i.e. [Z]=-3.000). Working in "log" space, as Arimoto & Yoshii (1987) did, would yield a mean luminosity-weighted metallicity of < [Z] > =((-3.0$\times$9)+(0.0$\times$1))/10.0=-2.70, whereas strict adherence to equation 2 would lead to <Z>=((0.00002$\times$9)+(0.02$\times$1))/10.0= 0.002018, or [<Z>]=-0.996. In other words, equation 2 would imply a mean metallicity 50 times higher than Arimoto & Yoshii (1987) would find.

It is clear from the work of Worthey (1994,1995) that using equation 2 is more representative of "reality" as most spectral indices scale much closer to Z (i.e. number of absorbers) than log Z. Of course, modelling the actual absorption features (e.g. Mg$_2$ line index) is the full and proper method for assigning metallicities, as the dependence upon Z is neither perfectly linear nor logarithmic, but for our purposes working with the linear regime for our "photometric" metallicity determinations is more than adequate.

$$<Z>_V = \left[ \sum_{i=0}^{t_{GW}/\Delta t} \sum_{j=1}^{n_{G-i}} \int_{m_{j,i,\ell}}^{m_{j,i,u}} \frac{\phi(m_{j,i})}{m_{j,i}} dm \, \psi(t_i) \, \Delta t \cdot L^{\Delta\lambda}_{j,G-i,Z(t_i)} Z(t_i) \right] \Big/ F^V(t_G) \quad (2)$$

It is important to be aware that it is common practice to only present a mass-weighted metallicity (e.g. Matteucci 1994; Bressan, Chiosi & Fagotto 1994; Elbaz, Arnaud & Vangioni-Flam 1995), which may overestimate the actual observed metallicity by $\gtrsim$ 0.2 dex, even for massive galaxies (e.g. Yoshii & Arimoto 1987; Gibson 1995). Comparison with the observed metallicity-luminosity relation should be made using the luminosity-weighted (i.e. equation 2), rather than the mass-weighted, metallicity.

### 2.3 The bimodal star formation model

The general bimodal star formation model for the evolution of elliptical galaxies was introduced by Arnaud et al.

(1992), and described more quantitatively by Elbaz, Arnaud & Vangioni-Flam (1995). As this particular formalism for "bimodal" evolution is a popular one (e.g. its use in modeling the ultraluminous IRAS 10214+4724 – Elbaz et al. 1992), we shall use their nomenclature, and refer the reader to their papers for any details not apparent in the description which follows.

The evolution of an elliptical is considered in four distinct phases. Phases I and III are the two in which star formation occurs. The first phase is strictly a high mass "burst" mode in which a Salpeter (1955) IMF is assumed, but with a high low-mass cut-off of $m_\ell = 3.0$ M$_\odot$ (and normal upper mass of 100 M$_\odot$), similar to that seen in certain starburst phenomena (Charlot et al. 1993). The star formation is intense, and short-lived ($\sim 15 \rightarrow 25$ Myrs for the initial masses considered below), due to the ensuing large SNe Type II rate, which subsequently drives a galactic wind for $\sim 30 \rightarrow 45$ Myrs (i.e. the so-called Phase II), thereby halting temporarily further star formation. Finally, the system enters a longer-lived ($\sim 1.5 \rightarrow 2.5$ Gyrs) phase (i.e. Phase III) of milder star formation, again with a Salpeter (1955), but this time over the full range of masses observed today (i.e. $\sim 0.1 \rightarrow 100$ M$_\odot$). Phase III star formation terminates with the onset of a second, smaller, galactic wind, once the thermal energy in the ISM exceeds that of its binding energy (Larson 1974). The subsequent, quiescent, Phase IV lasts until the current epoch.

Star formation during Phase I is exponential in time ($t$ in Gyr):

$$\psi_{\rm I}(t) = \nu_1\, e^{-t/\tau} M_{\rm g}(0) \qquad [{\rm M}_\odot/{\rm Gyr}] \qquad (3)$$

where $M_{\rm g}(0)$ is the initial gas mass in M$_\odot$; the astration parameter in Phase I is taken to be $\nu_1 = 40$ Gyr$^{-1}$; and, $\psi_{\rm I}$ decays exponentially with $\tau = 0.05$ Gyr. The gas mass and mass of metals Z in the gas during Phase I evolve as

$$\frac{{\rm d}M_{\rm g}(t)}{{\rm d}t} = -\psi_{\rm I}(t) + \int_{\max[m_t, 3.0]}^{100.0} \phi(m)\psi(t-\tau_m)R(m){\rm d}m \qquad (4)$$

and

$$\frac{{\rm d}M_{\rm Z}(t)}{{\rm d}t} = -{\rm Z}(t)\psi_{\rm I}(t) + \int_{\max[m_t, 3.0]}^{100.0} \frac{\phi(m)}{m}\psi(t-\tau_m) m_{{\rm Z},m}^{\rm ej}{\rm d}m \qquad (5)$$

respectively. The second terms on the right-hand sides of equations 4 and 5 represent the ejection rate of gas and metals from dying stars. The Schaller et al. (1992) Z=0.02 stellar lifetimes $\tau_m$ were adopted. Parallel models run with the Schaller et al. (1992) Z=0.001 lifetimes were negligibly different, corroborating the conclusion of Gibson (1995) that the lack of full stellar lifetime metallicity-dependence is not a major source of uncertainty. The yields $m_{{\rm Z},m}^{\rm ej}$ come from the recent metallicity-dependent compilation of Woosley & Weaver (1995), and we use the "mass in/mass out" formalism of Timmes, Woosley & Weaver (1995) (as opposed to Talbot & Arnett's (1973) matrix form) for solving equation 5. $R(m) \equiv 1 - w_m/m$ represents the fractional mass of a star of initial mass $m$ and remnant mass $w_m$ (from Prantzos, Cassé & Vangioni-Flam 1993) which is ejected back into the ISM after its lifetime $\tau_m$. $\phi(m) \propto m^{-x}$ is in the IMF, by mass, and $x = 1.35$ is the classical Salpeter (1955) slope used here. We stress that the ultimate conclusions of Section 3, regarding the bimodal star formation model, are not dependent upon the adopted stellar lifetimes, remnant masses, or stellar yields.

In Phase II, star formation is taken to cease for the duration of the galactic wind, and the total mass in the system $M_{\rm G}$ is assumed to evolve (i.e. decrease due to the steady wind outflow) as

$$\frac{{\rm d}M_{\rm G}(t)}{{\rm d}t} = -\alpha\, M_{\rm G}(t) \qquad [{\rm M}_\odot/{\rm Gyr}] \qquad (6)$$

where $M_{\rm G} = M_{\rm g}(0) - M_{\rm ej}$, and $M_{\rm ej}$ is the total mass ejected up to time $t$; $\alpha^{-1}$ is the characteristic exponential timescale for mass loss – $\alpha \equiv 18$ Gyr$-1$. The mass of metals in the remaining gas at each step is also adjusted accordingly.

The outflow stops once the gas has cooled to a level below that of the gas' binding energy, and the Phase III star formation regime is entered, with

$$\psi_{\rm III}(t) = \nu_2\, M_{\rm g}(t) \qquad [{\rm M}_\odot/{\rm Gyr}] \qquad (7)$$

In other words, star formation is taken to be proportional to the available gas mass, but at a significantly reduced level to that encountered in Phase I. The astration parameter $\nu_2 = 2$ Gyr$^{-1}$.

The evolution of the gas and metal mass in Phase III parallels that of equations 4 and 5, except that $\psi_{\rm I}$ is obviously replaced with $\psi_{\rm III}$, and the lower limits on the integrals are replaced with $\max[m_t, 0.1]$. During Phase III there is also the additional complication of gas being returned to the system from stars which were born during Phase I, but whose ejecta did not partake of the outflow of Phase II. Thus for the first $\sim 0.3$ Gyr (i.e. the lifetime of a 3 M$_\odot$ star minus the duration of the outflow) of Phase III there is an additional "ejecta" term for the dying lower-mass Phase I stars.

To ensure compatibility with the Elbaz, Arnaud & Vangioni-Flam (1995) results, we have set explicitly the Phase I, II, and III demarcation (e.g. wind initiation/completion) times given in their Table 2b, as opposed to recalculating explicitly the gaseous thermal energy evolution. While the numerical examples highlighted in this paper were selected on the basis of their being the favoured Elbaz, Arnaud & Vangioni-Flam (1995) models, the models hold, in general, for most permutations of this simple bimodal scenario. Let us now turn to the specific photo-chemical predictions for the model elliptical galaxies.

## 3  DISCUSSION

We ran models identical to Elbaz, Arnaud & Vangioni-Flam's (1995) preferred grid (Models A→D in their Table 2a). The initial gas masses were $5 \times 10^9$, $5 \times 10^{10}$, $5 \times 10^{11}$, and $5 \times 10^{12}$ M$_\odot$. No dark matter halos were considered for these particular models, although their Models E→H, with said halos, show this to be an unimportant consideration for what follows.

Figure 4 shows the evolution of each galaxy's gas mass as a function of time. The gas mass decreases during Phase I due to star formation, and continues to decrease in an overall sense during Phase II (due to the SNe Type II-driven outflow). The vertical line marks the onset of Phase III. We draw attention to the onset of Phase III because only stars formed during this phase contribute directly to the observed photo-chemical properties at the current epoch. The gas mass increases slightly for a few tenths of a Gigayear after the initiation of Phase III due to to gas returned by dying stars formed in Phase I. Once the last of these ($m = 3$ $M_\odot$) have died, the gas mass decreases drastically once again due to the star formation governed by equation 7. The behaviour of the gas mass *fraction* evolution is very similar for each of the models concerned. Figure 4 is directly comparable with the ISM curve in Figure 3 of Elbaz, Arnaud & Vangioni-Flam (1995), although it should be noted that we have normalised to the initial gas mass whereas they normalise to the current galactic mass. The single curve that they show ($M_g(0) = 5 \times 10^{11}$ $M_\odot$) is indistinguishable from that shown in our Figure 4, once this difference in normalisation is accounted for.

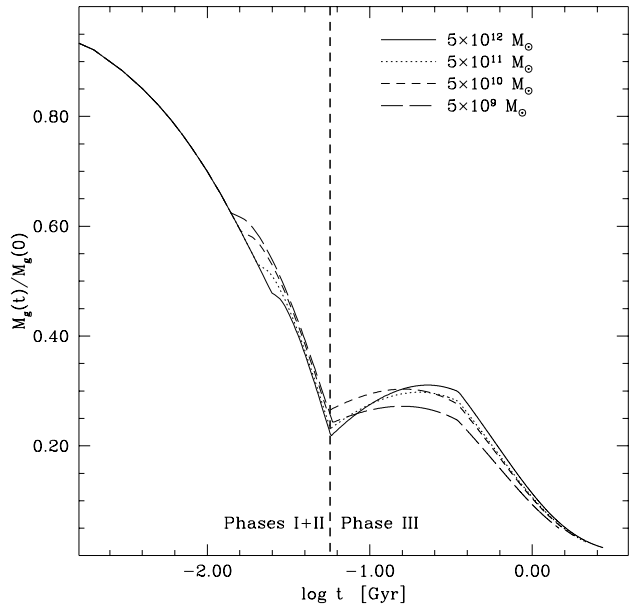

**Figure 4.** Evolution of the gas mass fraction of the ISM for the bimodal star formation models of Section 2.3. Only stars formed during Phase III are still alive at the current epoch.

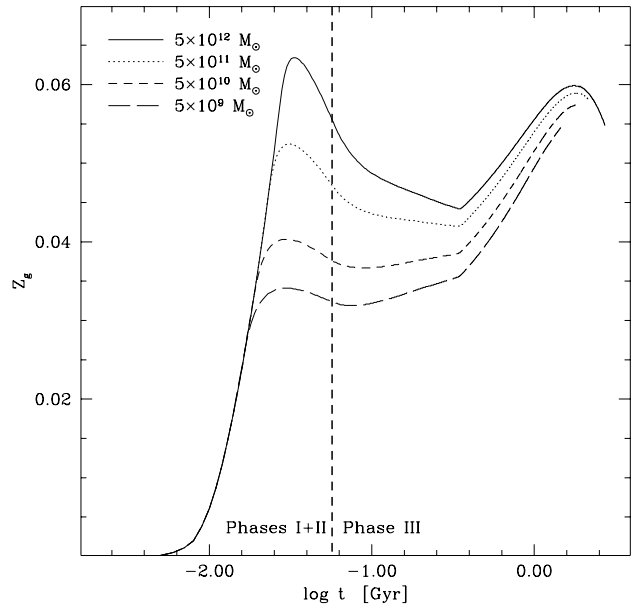

**Figure 5.** Evolution of the ISM metallicity $Z_g$ for the bimodal star formation models of Section 2.3. Only stars formed during Phase III are still alive at the current epoch.

Of more interest are the results shown in Figure 5. Here we see the evolution of the global metallicity of the ISM for the four different initial masses. The rapid rise of Z during Phase I is traced to the high initial star formation rates, and more importantly, the imposed high-mass mode of the IMF. The ISM metallicity passes the solar value in $\sim 14$ Myrs.

The metallicity during the initial few tenths of a Gigayear in Phase III increases very slowly, or even decreases, from its initial value, due to dilution effects from lower mass stars formed at lower Z in Phase I dying during the early portion of Phase III. This effect is enough to compensate for the increased metal production from the high-mass stars now dying in Phase III which themselves were formed in the pre-enriched (from Phase I stars) Phase III ISM. After $t \approx 0.3$ Gyr, the global Z increases dramatically for $\sim 1$ Gyr due to the cessation of this Phase I dilution influence.

Figure 6 shows the metallicity evolution as measured by [Z] and [Fe], for the $M_g(0) = 5 \times 10^{11}$ $M_\odot$ model. Allowing for the different normalisation, the [Fe] curve is indistinguishable from that shown in Figure 4 of Elbaz, Arnaud & Vangioni-Flam (1995) (note: they normalise to the total initial mass). We note, further, that [Fe]≈[Fe/H] for these models. [Fe/H] *roughly* tracks the global metallicity [Z] in Phase III, although offset by typically $\sim 0.2 \rightarrow 0.3$ dex (although always super-solar in Phase III, regardless of metallicity "definition"). See also the comment at the end of this section regarding the SNe Ia formalism adopted by Elbaz, Arnaud & Vangioni-Flam (1995) which may mean that the iron abundance of Figure 6 is actually an underestimate.

In this work we generally refer to the evolution of the global metallicity Z, as opposed to the metallicity denoted by [Fe/H], simply because the observed "[Fe/H]" tabulated in Terlevich et al. (1981), for example, are actually closer to [Z] than to [Fe/H], per se (Terlevich 1995), as they are indirectly assigned from a combination of spectroscopic indicators ($Mg_2$, in this case) and photometric calibrators (e.g. Barbuy 1994, and references therein). This is confirmed by González & Gorgas (1995). Obviously, neither [Z] nor [Fe/H] are ideal, but as the current version of our code is only "photometric" in nature (i.e. no spectral line indices), we chose the best compromise and compared our V-band luminosity-weighted metallicity [<Z>]$_V$ against the observations shown in Figure 7. Elbaz, Arnaud & Vangioni-Flam (1995) only report the mass-weighted [Fe/H] metallicity for their models, so this difference should be kept in mind, although we stress that any differences which might crop up due to this [Z] versus [Fe/H] metallicity "definition" do not affect the conclusions of this paper.

Recalling that only stars formed during the "normal" IMF mode of Phase III are still alive today to contribute to the elliptical's luminosity (and hence photo-chemical properties), we can anticipate immediately from Figure 5 that there is going to be a problem with their bimodal star formation scenario. Specifically, each of the Phase III curves only span $\lesssim$ 0.2 dex in [Z], whereas observations show a metallicity spread of $\sim$ 3.5 dex (e.g. Minniti et al. 1995, and references therein). Not only is their implied metallicity distribution virtually non-existent, but there are no stars with $Z \lesssim (2 \rightarrow 3) Z_\odot$ (again, at odds with the inferred low-metallicity components in ellipticals – e.g. Rose 1985). We conclude that just on the basis of this rough "chemical argument" there would seem to be concern for the bimodal star formation scenario. Still, let us play devil's advocate and claim that since there is only minimal *direct* evidence for metallicity spreads in ellipticals (e.g. a dispersion $\gtrsim$ 2 dex is seen in the giants in the nearby dwarf ellipticals M32 and NGC 147 – Davidge & Jones 1992; Davidge 1994 – the bulk of the evidence is *indirect* in nature – i.e. based upon spiral bulge stars – e.g. Minniti et al. 1995), this chemical argument alone is not enough for dismissing the simple bimodal model. Unfortunately, the problems become even more acute when we look at their predicted present-day colour-metallicity-luminosity relations – i.e. direct observational tests which *do* hold for local ellipticals.

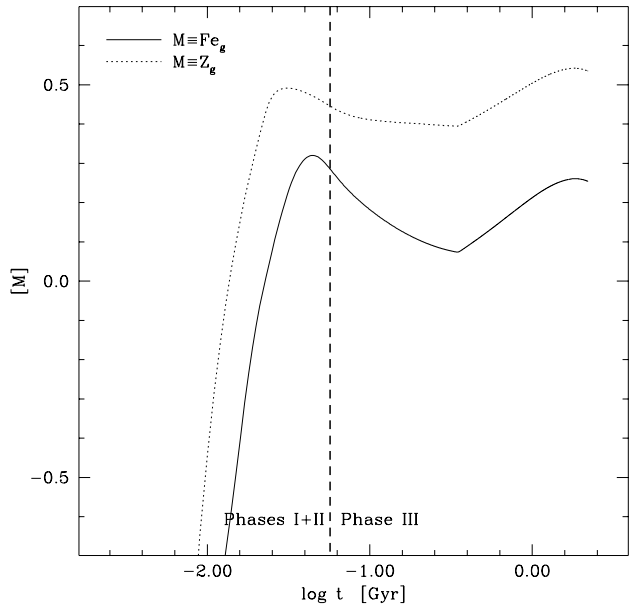

**Figure 6.** Evolution of the global ISM metallicity [Z] and its iron content [Fe] for the bimodal star formation model of initial mass $5 \times 10^{11}$ $M_\odot$. The [Fe] curve matches that of Elbaz, Arnaud & Vangioni-Flam's(1995) Figure 4.

The framework adopted by Elbaz, Arnaud & Vangioni-Flam (1995) (and most of the wind model papers referenced thus far) – coeval galaxies with the present-day metallicity-luminosity relation driven by the cessation of star formation due to the galactic winds – is a typical, and elegantly simple, one in which to work, but it should be stressed that this may not be a unique interpretation. For example, Worthey, Trager & Faber (1995) speculate that the observed relations actually reflect a luminosity-weighted age-metallicity relation with the most metal-rich ellipticals being $\sim$ 2 Gyr old (i.e. later phases of star formation taking place in pre-enriched gas from an older epoch of star formation), and the most metal-poor $\sim$ 15 Gyr (i.e. little or no late-time star formation). These Worthey, Trager & Faber (1995) ages are based on $C_24668$ line strengths and an average of Balmer indices, *but* as they themselves stress, it only takes a "frosting" of a fraction of a percent of a very young population on top of a basically old population to impact dramatically upon the Balmer line strengths, making a galaxy appear much younger than a mass-weighted mean age (Section 3 of Worthey, Trager & Faber 1995). Obviously, adopting any sort of "mean" age or "mean" metallicity will mask, at some level, the true complexity of the star formation process in ellipticals, a model limitation which should always be kept in mind.

Figure 7 shows the predicted metallicity-luminosity relation for the grid of models described above (the two dotted lines represent the relevant photo-chemical predictions derived from the two isochrone sources – in this plane, the Worthey (1994) and Bertelli et al. (1994) predictions are virtually indistinguishable). The solid curve represents the template of models outlined in Gibson & Matteucci (1995)[*]. As we could anticipate from Figure 5, the Elbaz, Arnaud & Vangioni-Flam (1995) present-day luminosity-weighted metallicities are considerably greater than what is seen in nearby ellipticals – $<[Z]>_V$ is discrepant with the observations by $\sim 0.2 \rightarrow 1.0$ dex, the predictions getting progressively worse for lower masses. Again, as anticipated from Figure 7, it is not difficult to trace the problem to the complete lack of low-metallicity stars at $t_G = 12$ Gyr. *The high mass star formation mode of Phase I may be an efficient mechanism for ejecting vast quantities of O and Fe to the cluster ICM, but its unavoidable enrichment of the Phase III ISM means that no sub-solar metallicity stars remain today.* This is at odds with the direct and indirect arguments for sub-solar metallicity components in nearby ellipticals, put forth by Rose (1985), Arimoto & Yoshii (1987), Hardy et al. (1994), and Minniti et al. (1995), amongst others.

---

[*] The Gibson & Matteucci (1995) template adopts a single, somewhat flatter-than-Salpeter (1955) IMF ($x \approx 1$), with $m_\ell = 0.2$ $M_\odot$ and $m_{\rm u} = 65$ $M_\odot$; Worthey's (1994) isochrones were assumed. Early galactic wind times (i.e. $t_{\rm GW} \lesssim 0.1$ Gyr), similar to those favoured by Bressan, Chiosi & Fagotto (1994) (whose colour-luminosity predictions are similar to the Gibson & Matteucci (1995) grid's in Figure 8), were found, despite our conservative approach to pre-SN stellar wind energy deposition to the ISM (Gibson 1994a), and their extreme view that SNe are only negligible contributors in setting $t_{\rm GW}$. Indeed, despite the analysis of Gibson (1994a), Bressan, Chiosi & Tantalo (1995) and Tantalo et al. (1995) remain adamant that if the colour-metallicity relation of ellipticals is to be understood, stellar wind energy must dominate. How then can the Gibson & Matteucci (1995) photo-chemical predictions (and, in particular, $t_{\rm GW}$) be comparable with those of Bressan, Chiosi & Fagotto's (1994)? The answer lies in the treatment of SN remnant dynamics and thermal energetics. Bressan, Chiosi & Fagotto's (1994) treatment (and indeed, that of Arimoto & Yoshii 1987; Matteucci & Tornambe 1987; Angeletti & Giannone 1990; to name but a few) is the classic "evolution in isolation" formalism in which remnants are assumed to expand and cool in complete isolation *ad infinitum*, resulting in the inefficient transfer of thermal energy to the ISM. This fundamental assumption inherent in most previous galactic wind models is incorrect, and does not reflect the complexity of reality – in reality, as recognised by Larson (1974), Dekel & Silk (1986), Babul & Rees (1992), Gibson (1994b), and Nath & Chiba (1995), shells quickly come into contact and overlap with neighbouring, expanding shells, affecting strongly both their dynamics and energetics. Subsequent SNe explosions occur in the resultant, expanding, rarefied bubble – i.e. "superbubble" (Tomisaka 1992) – with very little adiabatic or radiative cooling energy loss. In reality then, the SNe energy transfer to the ISM is very efficient, and early galactic wind times are a natural outcome; there is no need to resort to what we feel are physically implausible assumptions regarding the stellar wind energy, as was done by Bressan, Chiosi & Fagotto (1994), Bressan, Chiosi & Tantalo (1995), and Tantalo et al. (1995), in order to fit the galactic colour-luminosity relation. The exact formalism adopted for the SNe energetics is given by Model $B'_3$ of Gibson (1994b), whereas that adopted by Bressan, Chiosi & Fagotto (1994) would correspond to Gibson's (1994b) Model $A_0$. Further details can be found in Gibson (1994b), Gibson & Matteucci (1995), and Gibson (1995). The latter work addresses the Bressan, Chiosi & Fagotto (1994) paper specifically.

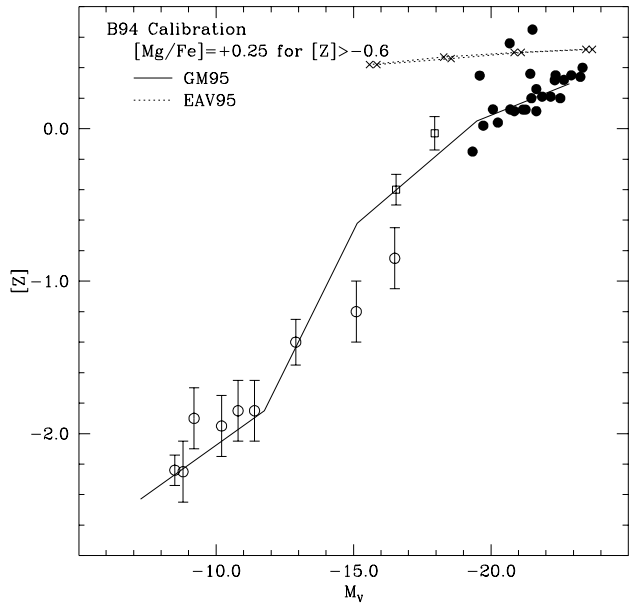

**Figure 7.** Observed metallicity-luminosity relation for dwarf (open circles: Smith 1985) and "normal" (open squares: Sil'chenko 1994 – filled circles: Terlevich et al. 1981) ellipticals. The [Mg/Fe]=+0.25, Mg$_2$-[Z] calibration of Barbuy (1994) has been adopted for the normal/giant ellipticals. The dotted lines represent the predictions for the bimodal star formation/IMF models of Elbaz, Arnaud & Vangioni-Flam (1995) (the two predictions come from the Worthey (1994) and Bertelli et al. (1994) isochrones, and are virtually indistinguishable), while the solid curve shows the template models of Gibson & Matteucci (1995).

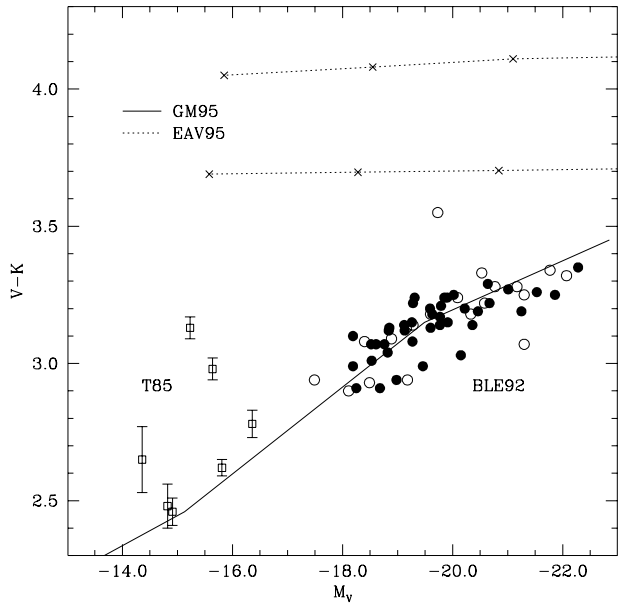

**Figure 8.** V-K colour-luminosity relation for Virgo (open circles) and Coma (filled circles) cluster ellipticals and lenticulars (Bower, Lucey & Ellis 1992). A Virgo distance modulus of $(V - M_V)_0 = 31.54$ is assumed, and a shift of $\Delta(V - M_V)_0 = 3.58$ has been applied to the Coma sample. Data for the dwarfs from Thuan (1985). The dotted lines represent the predictions for the bimodal star formation/IMF models of Elbaz, Arnaud & Vangioni-Flam (1995) (upper line implied use of the Worthey (1994) isochrones, while the lower line used the Bertelli et al. (1994) isochrones), and the solid curve shows the template models of Gibson & Matteucci (1995).

The problems are equally apparent in the colour-luminosity plane. In Figure 8 we show once again, the predicted V-K for the Elbaz, Arnaud & Vangioni-Flam (1995) bimodal star formation models (dotted lines, where the upper line shows the colours predicted from the Worthey (1994) isochrones, and the lower line, the Bertelli et al. (1994) isochrones), as well as those for the proposed template of Gibson & Matteucci (1995) (solid curve). Again, the Gibson & Matteucci (1995) models use the more conventional Arimoto & Yoshii (1987) IMF, and satisfy the observational constraints of both the ICM *and* the underlying stellar populations. The Elbaz, Arnaud & Vangioni-Flam (1995) models are far too red – by $\sim 0.8 \rightarrow 1.4$ mag, when using the Worthey (1994) isochrones, and by $\sim 0.4 \rightarrow 1.0$ mag, even when using the bluer Bertelli et al. (1994) isochrones. Recalling that there are no Z$\lesssim 0.03$ stars left at the current epoch in their models, this should not be too surprising. Table 1 shows us that the bluest component in these CSPs would then only have V-K$\approx$3.9 (Worthey) or $\approx$3.6 (Bertelli et al. ). We remind the reader that the Gibson & Matteucci (1995) template was tuned to fit the observed the colour-metallicity-luminosity relations of Figures 7 and 8 via use of the Worthey (1994) isochrones. Adopting the Bertelli et al. (1994) isochrones, in their place, would necessitate the change of some of the specific details (e.g. star formation efficiency $\nu$, galactic wind epoch $t_{GW}$ – see also Gibson 1995) in order to recover the observed relations, but the general conclusion of the current study would remain unchanged.

We mention in passing that the preferred grid of models from Elbaz, Arnaud & Vangioni-Flam (1995) (Models A$\rightarrow$ D) assumed a Type Ia SN rate formalism which is un-evolving in time, given roughly by the present-day observed rates (Turatto, Cappellaro & Benetti 1994). Considering that adopting more physically significant models for SNe Ia evolution (e.g. Figure 1 of Matteucci & Tornambe 1987) implies that said rate in the initial $\sim$ 2 Gyrs of star formation should be $\sim$ 20 times greater than the present-day value (for the singly-degenerate Type Ia SNe progenitor model of Whelan & Iben 1973), there is the potential to have underestimated the ISM iron content seen in Phase III of Figure 6. This would have the tendency to drive down their resultant stellar [Mg/Fe] predictions (which currently are consistent with the observations of Worthey, Faber & González 1992), and increase their stellar [Fe/H], thereby worsening the already poor colour-metallicity-luminosity predictions shown in Figures 7 and 8.

## 4  SUMMARY

We re-iterate the conclusions drawn in Section 3 – the simplest bimodal IMF/star formation models for elliptical galaxies, with galactic winds (i.e. those of Arnaud et al. (1992) and Elbaz, Arnaud & Vangioni-Flam (1995)), may satisfy the ICM observational constraints, but only at the expense of the resultant predicted present-day photo-chemical properties for the underlying ellipticals. Specifically, the predicted V-K colours are $\sim 0.5 \rightarrow 1.4$ magnitudes too red, the luminosity-weighted metallicities $\sim 0.2 \rightarrow 1.0$ dex too high, and the intrinsic stellar metallicity dispersion is virtually non-existent (i.e. $\lesssim 0.2$ dex, versus the $\gtrsim 3.5$ dex implied observationally). Renzini et al. (1993) have also criticised the simple bimodal models, albeit on more general terms than those presented in this paper. We conclude that the more conventional models with flatter-than-Salpeter (1955) IMFs and efficient SN thermal energy transfer to the ICM, and which use the photometric techniques outlined in this paper, are still the preferred scenario for elliptical galaxies with galactic winds (Gibson & Matteucci 1995), in our opinion.

## ACKNOWLEDGEMENTS

I wish to thank Guy Worthey and Paolo Bertelli for a number of useful correspondences. The hospitality shown by the Astrophysics Department at Oxford during my extended "sabbatical" is gratefully acknowledged, as is the financial assistance of NSERC, through grants to Paul Hickson and Kristine Hensel. Finally, the anonymous referee is thanked for improving significantly the paper's content and presentation, as well as for drawing our attention to the Tantalo et al. preprint.